# A Medium Sized Schwarzschild-Couder Cherenkov Telescope Mechanical Design Proposed for the Cherenkov Telescope Array

K. Byrum[1], B. Humensky[2], W. Benbow[3], R. Cameron[4], S. Criswell[5], M. Errando[2], V. Guarino[1], P. Kaaret[6], D. Kieda[7], R. Mukherjee[2], D. Naumann[8], D. Nieto[2], R. Northrop[9], A. Okumura[10,11,12], E. Roache[5], J. Rousselle[13], S. Schlenstedt[8], R. Sternberger[8], V. Vassiliev[13], S. Wakely[9], H. Zhao[1] for the *CTA Consortium

[1]*Argonne National Laboratory, 9700 S. Cass Ave. High Energy Physics Div., Lemont, IL 60439,*
[2]*Columbia University, Physics Department, Department of Physics and Astronomy, Barnard College, New York, NY 10027,* [3]*Harvard-Smithsonian Center for Astrophysics, Cambridge, MA 02138,* [4]*SLAC National Accelerator Laboratory, 2575 Sand Hill Rd, Menlo Park, CA 94025,* [5]*Fred Lawrence Whipple Observatory, 670 Mount Hopkins Rd, Amado, AZ 85645,* [6]*University of Iowa, Department of Physics and Astronomy, Iowa City, IA 52242,* [7]*University of Utah, Physics Department, Salt Lake City, UT 84112,* [8]*Deutsches Elektronen-Synchrotron(DESY), Platanenallee 6, 15738 Zeuthen, Germany,* [9]*University of Chicago, Enrico Fermi Institute, Chicago, IL 60637,* [10]*Solar-Terrestrial Environ. Lab, Nagoya Univ, Furo-cho, Chikusakku, Nagoya, Aichi 464-8601, Japan,* [11]*University of Leicester, Department of Physics and Astronomy, Leicester, LEI 7RH, UK,* [12]*Max-Planck-Institut fur Kernphysik, P.O. Box 103980, D 69029 Heidelberg, Germany,* [13]*University of California, Los Angeles, Department of Physics and Astronomy, LA, CA 90095*
E-mail: byrum@anl.gov

The Cherenkov Telescope Array (CTA) is an international next-generation ground-based gamma-ray observatory. CTA will be implemented as southern and northern hemisphere arrays of tens of small, medium and large-sized imaging Cherenkov telescopes with the goal of improving the sensitivity over the current-generation experiments by an order of magnitude. CTA will provide energy coverage from ~20 GeV to more than 300 TeV. The Schwarzschild-Couder (SC) medium size (9.5m) telescopes will feature a novel aplanatic two-mirror optical design capable of accommodating a wide field-of-view with significantly improved angular resolution as compared to the traditional Davies-Cotton optical design. A full-scale prototype SC medium size telescope structure has been designed and will be constructed at the Fred Lawrence Whipple Observatory in southern Arizona during the fall of 2015. This report will concentrate on the novel features of the design.

[2]Presenter
*Full consortium author list at http://cta-observatory.org

## 1. Introduction

The Cherenkov Telescope Array (CTA) [1] is an international next-generation ground-based gamma-ray observatory. CTA will be implemented as southern and northern hemisphere arrays of tens of small, medium and large-sized imaging Cherenkov telescopes with the goal of improving the sensitivity over the current-generation experiments like HESS, MAGIC and VERITAS by an order of magnitude. CTA will provide energy coverage from ~20 GeV to more than 300 TeV. Davies-Cotton (DC) single-mirror telescope designs or parabolic single-mirror designs (both being segmented mirrors) have traditionally been used for Cherenkov telescopes and single-mirror designs are the baseline for the CTA large and medium sized telescopes. However, they are dominated by the cost of the camera, which is based on classical 1in to 2in photo-multipliers paving a large focal plane area, hence making single-mirror designs not ideal for wide field-of-view (FOV). An alternative to the traditional DC telescope is a Schwarzschild-Couder (SC) telescope which features a novel aplanatic two-mirror optical design that fully corrects for spherical and comatic aberrations [2] while capable of accommodating a wide FOV with significantly improved angular resolution as compared to the traditional Davies-Cotton optical design. SC designs are being explored for CTA with the goal of complementing the traditional DC designs with additional SC telescopes. In this paper, we describe the novel features of a full-scale SC medium-sized telescope (SC-MST) that has been designed. The prototype SC telescope (pSCT) structure will be constructed at the Fred Lawrence Whipple Observatory in southern Arizona during the fall of 2015.

## 2. The pSCT Optical Support Structure Design

The pSCT is a two-mirror telescope with a counterweight structure. The Optical Support Structure (OSS) which supports the mirrors and camera will be mounted on a positioner composed of two main parts: the head and yoke, and the tower. The positioner provides the motion about the elevation and azimuth axes. The positioner is the baselined DESY Davies-Cotton medium-sized telescope (DC-MST) design. The only difference between the positioner used for the DC-MST and the positioner used for the pSCT is the height of the tower for the pSCT design which is shorter. Figure 1 shows the telescope in various orientations.

The pSCT is composed of primary and secondary mirrors and of a camera. The performance of the pSCT is dependent on the relative deformation of each of these components to each other. It is important to keep the axes of revolution for the primary, secondary and focal planes parallel to the optical axis of the telescope, as well as to maintain the relative distance between surfaces.

Many different geometries of the SC-MST have been examined using simple beam/shell element Finite Element Analysis (FEA) models. The purpose of these analyses has been to understand the deformations and how they influence the Point Spread Function (PSF) so that a basic geometry of the structure can be determined and carried forward. From this extensive analysis, the current design was developed to take advantage of previous work on the DC-MST dish and to allow for easy integration with the DESY positioner.

The design has been driven by the following criteria:
- PSF performance
- Minimization of shadowing

- Ease of fabrication of parts
- Ease of shipping and assembly on-site

The requirement for the PSF performance is less than 3 arcmin at 3 degree field angle. Simple beam and shell element models were created to evaluate the PSF performance based on design implementation. The mirrors were modeled as a single point on the OSS that corresponded with the center of a triangle that would support the hexapod which in turn would support the mirrors. In all of the designs, the camera weight was 700 kg. The mass of mirror segments with a hexapod positioning system was 42 kg each. The mass of the secondary structure and mirrors was 7 tons. The average mass of the primary structure with the mirrors was 13 tons.

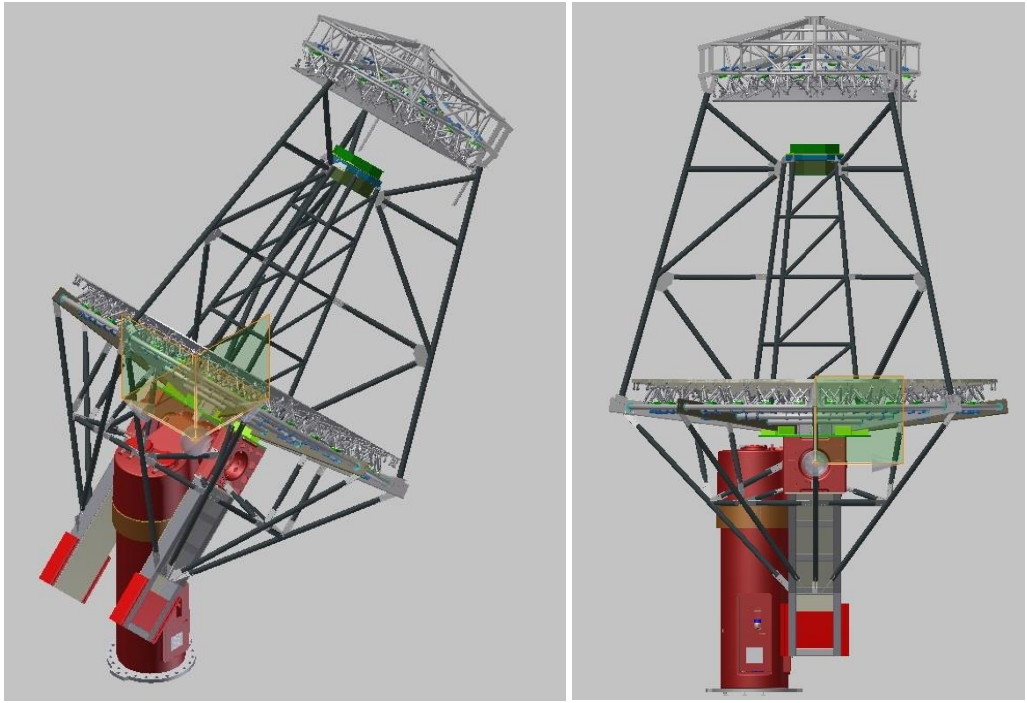

Figure 1. pSCT telescope design.

## 3.   Specifications

The SC-MST telescope is required to meet certain requirements that are defined by the CTA collaboration as well as requirements that are specific to the SC-MST. These are described in detail in the Technical Design Report [3]. We list some of the parameters most important for the pSCT structure design below.

### 3.1  Mechanical Parameters

| Parameter | Definition |
|---|---|
| Life expectancy (for bearings and structure) | 30 years |
| Eigenfrequency (whole structure) | $\geq 2.5$ Hz |
| Tracking accuracy | $\leq 1.2$ arcmin |
| Displacement of camera due to loading | $\leq$ ½ pixel ~ 19 mm |
| Angular misalignment of camera | not significant |
| Weight of mirrors including actuators | 35 kg/m² |

Table 1. Mechanical Parameters for the pSCT.

## 3.2 Optical Parameters

| Parameter | Definition |
|---|---|
| Structure Name | Schwarzschild-Couder (SC) |
| Focal Length (F) [m] | 5.5863 |
| Aperture [m] | 9.66 |
| Primary Radius max [m] | 4.8319 |
| Primary Radius min [m] | 2.1933 |
| Secondary Radius max [m] | 2.7083 |
| Secondary Radius min [m] | 0.3945 |
| Effective light collecting area /unvignetted [$m^2$] | 50.33 |
| Unvignetted Size [deg] | 3.50 |
| Effective light collecting area at FOV edge [$m^2$] | 47.73 |
| Vignetting at the FOV edge [%] | -5.17 |
| Primary project area [$m^2$] | 58.23 |
| Secondary project area [$m^2$] | 22.55 |
| Design FOV [deg] | 8.00 |
| Design FOV solid angle [$deg^2$] | 50.35 |
| Ideal PSF at the FOV edge [arcmin] | 3.81 |
| Mirror separation [m] | 3/2 x F |
| Separation of secondary to camera [m] | 1/3 x F |
| Shadowing | Less than 12% |

Table 2. Optical Parameters for the pSCT.

## 4. Description of Mechanical Design

The basic geometry with dimensions of the pSCT OSS is shown in Figure 2a. The layout of the OSS begins with the surface of the mirrors. The thickness of the mirror and the distance from the mirror to the support base is identical for every panel module. A typical panel module includes the mirror, Stewart Platform (SP) and support base. The design of the OSS began by placing the inner surface for the primary and secondary mirrors in their correct positions in the optical coordinate system. The steel structure of the OSS that supports each panel module was then located so that it provided support to the mirror assemblies using the dimensions shown in Figure 2b.

The design of the primary and counterweight structures is nearly identical to the design used for the DC-MST. The primary will be mounted to a main plate that is bolted to the yoke on the positioner and consists of 8 radial arms which are tied together by circular tubes. A very stiff primary structure is connected to the positioner, and the secondary and camera are supported by a frame fixed to the primary mirror structure. A counterweight is needed so that the entire structure is balanced about the elevation axis. The counterweight structure serves a secondary purpose of stiffening and minimizing deflections of the primary mirror. In the current design, a counterweight of 14 tons is needed. The secondary is supported by three arms that are deep trusses connected to the ouside diameter of the primary. Each of these trusses gain additional stiffness by being connected through the center truss that supports the camera.

This truss structure shadows the light onto the primary but does not shadow the light that travels from the primary to the secondary. The camera is supported from the inner diameter of the primary; therefore, there is no shadowing due to the camera and its supports. The primary and secondary structures will be made from simple steel tubes that are welded together at a workshop and then bolted together on-site. The goal is that no welding needs to occur on-site.

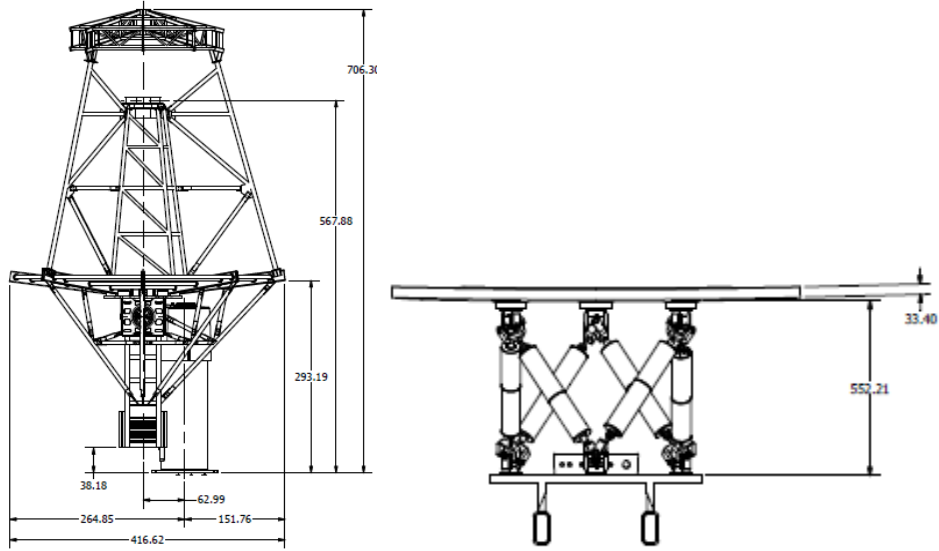

Figure 2. 2a-(left) Basic dimensions of the OSS (inches). 2b-(right) Assumed dimension of panel module.

**4.1 Camera Mounting**

The mounting and access of the camera [4] poses many challenges that are not present in a DC telescope where there is full access to the camera. The camera must be accessed through the structure that supports the secondary. In addition, access on the back side of the camera is limited by the structure supporting the camera.

The required PSF performance is premised on having a very stiff camera frame. The camera weight is anticipated to be 700kg which is relatively light. The camera will be mounted with the telescope horizontal and either lowered by crane or lifted from the ground using a portable lift. Once at the correct height the camera will be moved horizontally to slide into the existing framework and bolted into place. Figure 3 shows how the camera body is bolted to a stiff frame.

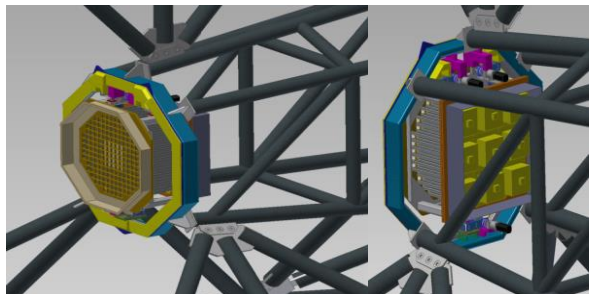

Figure 3. Camera mounting on the OSS.

**4.2 Mirror Mounting**

The mirrors for the pSCT [5] will be mounted to the hexapod SP which in turn will be mounted to a triangular base that is mounted to the curved beams of the primary and secondary. The triangular base has to be located on the OSS within 10mm and the SP has a large enough range of adjustment to accommodate this tolerance. The mounting of the mirrors is shown in Figures 4a and 4b.

Each panel module will also have an electronics box mounted to the center of the triangle base (Figure 4a) and alignment edge sensors (Figure 4b). The global and panel-to-panel alignment systems are described in detail in [6]. The entire mirror mounting assembly shown in Figure 4a weighs approximately 50kg. There are four different sets of mirror mount assemblies: Inner Primary; Outer Primary; Inner Secondary; Outer Secondary. Each panel module is very delicate and can only be handled by attaching lifting fixtures to the back surface of the mounting triangle. The entire assembly then must be lifted onto the telescope structure and mounted.

The base of each triangle assembly is a triangular aluminum casting. On the back of each triangle are two half-moon cutouts that will be cast to match the curved shape of the curved tube on the OSS that it will be mounted to. A casting is a cheap way to form this relatively large component and also achieve the unique shape of each half-moon cutout that is required for each of the four mirror assemblies.

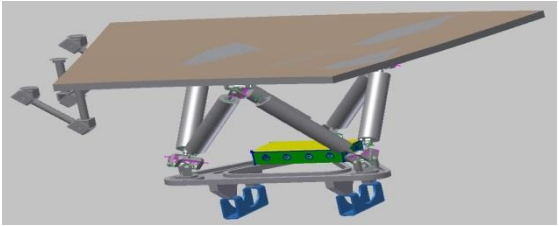

Figure 4a - Typical mirror mounted to stewart platform and base with edge sensors. The electronics box is shown in yellow and green in the figure.

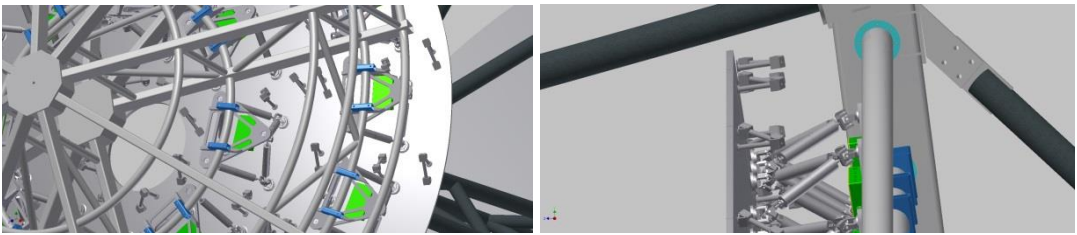

Figure 4b – (left) Mirror mounting on the secondary. The green triangle bases and edge sensors are shown. (right) Side view of mirror mounting to the primary.

The location and orientation of the mirrors would be set beforehand on the primary during fabrication. Since the mirror assemblies are relatively heavy and difficult to maneuver, especially high in the air on a man lift, it is important that the mounting points on the OSS be fixed and set the mirror assemblies to their final position. In Figure 4 above there are steel fixtures (in blue) that will be welded in the correct position and orientation to the OSS during fabrication.

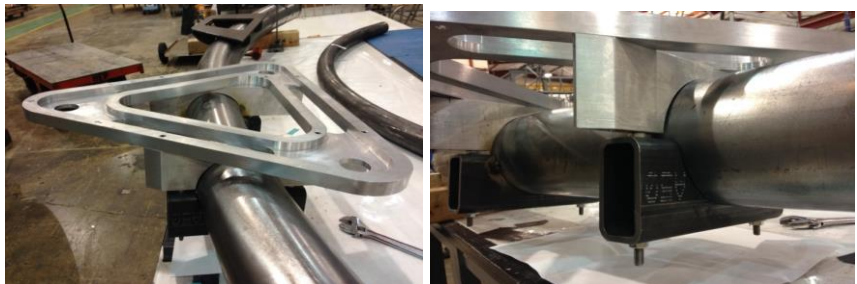

Figure 5 – Prototype mirror mounting fixture.

Figure 5 shows a prototype fixture that is welded to the curved tube and the cast aluminum triangle bolted to it to form a rigid assembly. The mirror assemblies then only have to be lowered onto the curved beams and bolted into place.

The mounting of the primary could occur in either the horizontal or vertical orientation. If the telescope is vertical then man lifts can easily be used to lift technicians underneath the primary to secure the mirrors. The mirrors would be lifted using special fixtures that would lift the mirror assemblies from the back and hold them in approximately the correct position. Figure 6 shows concepts for lifting fixtures for the mirror assemblies. The mounting of the mirrors to the secondary uses the same lifting fixtures with a similar procedure.

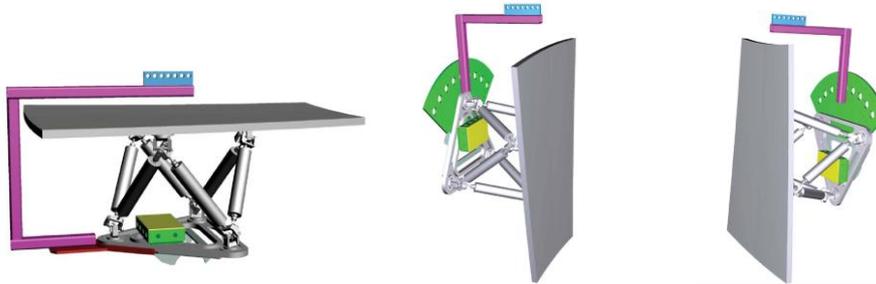

Figure 6 – Lifting fixtures (left) Horizontal mirror orientation, (Middle) Vertical orientation, (right) Inverted vertical orientation.

### 4.3 Baffles

Baffles are required for the primary and secondary telescope structure, both to reduce the night sky background scatter and to contain the reflected sunlight during the daytime for safety reasons. Figure 7 shows the baffles on the primary and secondary.

The secondary baffle is parallel to the optical axis and projects 1m past the front surface of the mirrors. The primary baffle has a 4 degree angle with respect to the optical axis and projects 1.5m past the front surface of the mirrors. Since the mirrors are already approximately 0.5m from the OSS the baffles have to extend an additional 0.5m to the OSS.

During mirror mounting the baffles would not be in place because they limit access to the OSS and would interfere with crane access. The baffles are ideally thin sheets of aluminum. The main challenges of the baffles is providing them with enough stiffness so that they maintain their shape and do not vibrate during wind gusts. Thin sheets of aluminum can gain stiffness by rolling ribs like 55 gallon drums have at their mid-section for stiffness. Stiffness could also be achieved by rolling small sections of tubes that are used to secure the baffles at the outer edges.

### 4.4 Park Position

During daylight hours the telescope will be parked facing north. The elevation angle in the park position will change throughout the year. The current park positions are -5 degrees, +20 degrees, and +45 degrees. The telescope will also be in the park position anytime the wind exceeds the maximum velocity of 50km/hr. In the park position the OSS, positioner and foundation must resist the maximum wind loads [3]. The torques on the elevation and azimuth axis will be resisted by stow pins. Figure 8 shows mechanism that will be used to secure the elevation axis. There is a large stow pin that protrudes from the head of the positioner and fits
in a hole in a fixture that is bolted to the back of the main plate of the OSS. The stow pin holes correspond to the park positions.

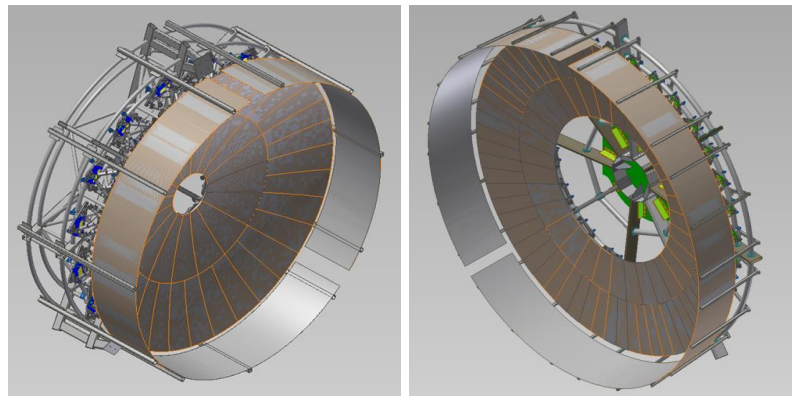

Figure 7 – (left) secondary Baffle; (right) primary Baffle.

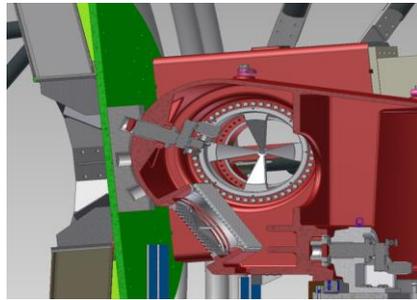

Figure 8 – Elevation axis lock for the park positions.

## 5.   Summary

In conclusion, the design for the pSCT OSS has been completed. The assembly plans have been developed and are currently being reviewed. The pSCT OSS is currently being fabricated by Walters Metal Fabrication and will be delivered and constructed at the Fred Lawrence Whipple Observatory in southern Arizona during the fall of 2015. Commissioning of the prototype will begin once the telescope and camera are completed in fall 2015.

## 6.   Acknowledgements

We gratefully acknowledge support from the agencies and organizations listed under Funding Agencies at this website: http://www.cta-observatory.org/. The development of the prototype SCT has been made possible by funding provided through the NSF-MRI program.